Review

# FAST: Its Scientific Achievements and Prospects

Lei Qian,[1,2] Rui Yao,[1,2] Jinghai Sun,[1,2] Jinlong Xu,[1,2] Zhichen Pan,[1,2] and Peng Jiang[1,2,*]

[1]National Astronomical Observatories, Chinese Academy of Sciences, Beijing 100101, China
[2]University of Chinese Academy of Sciences, Beijing 100049, China
*Correspondence: pjiang@nao.cas.cn
Received: August 11, 2020; Accepted: October 28, 2020; Published: November 25, 2020; https://doi.org/10.1016/j.xinn.2020.100053
© 2020

The Innovation

## GRAPHICAL ABSTRACT

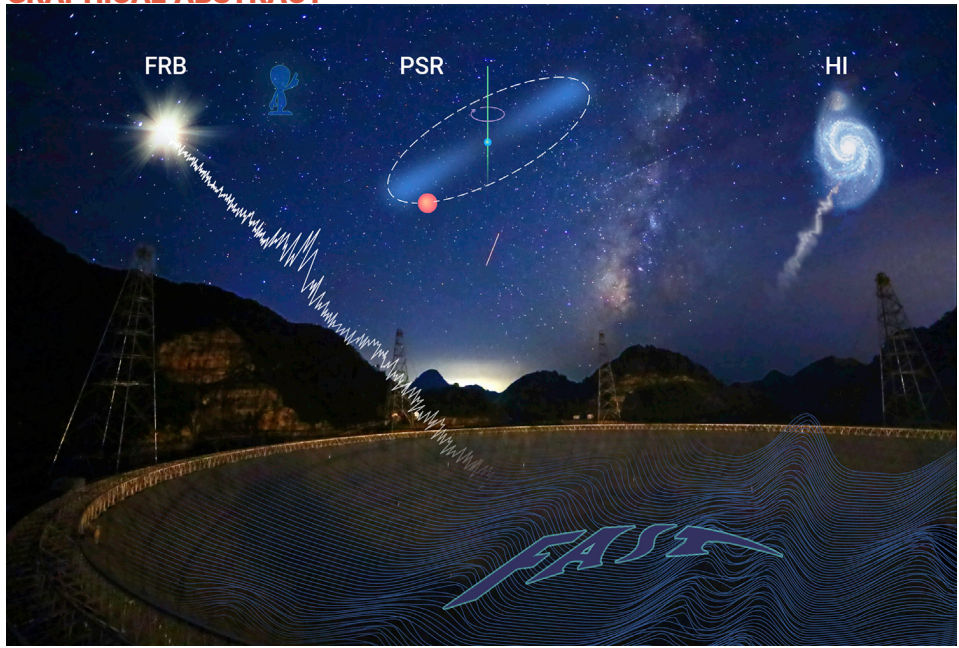

## CORRESPONDENCE

pjiang@nao.cas.cn

https://doi.org/10.1016/j.xinn.2020.100053

Received: August 11, 2020
Accepted: October 28, 2020
Published: November 25, 2020

www.cell.com/the-innovation
## PUBLIC SUMMARY

- FAST has found more than 240 pulsars, including pulsars in binary systems
- FAST has made great progress in the study of fast radio bursts
- The FAST team will call for proposals from the whole world in 2021

CellPress Partner Journal                                                                 www.cell.com/the-innovation

# FAST: Its Scientific Achievements and Prospects

Lei Qian,[1,2] Rui Yao,[1,2] Jinghai Sun,[1,2] Jinlong Xu,[1,2] Zhichen Pan,[1,2] and Peng Jiang[1,2,*]
[1]National Astronomical Observatories, Chinese Academy of Sciences, Beijing 100101, China
[2]University of Chinese Academy of Sciences, Beijing 100049, China
*Correspondence: pjiang@nao.cas.cn

FAST is the largest single-dish radio telescope in the world. The characteristics of FAST are presented and analyzed in the context of the parameter space to show how FAST science achievements are affected. We summarize the scientific achievements of FAST and discuss its future science based on the new parts of the parameter space that can be explored by FAST.

## Introduction

The study of the universe is an exploration of the parameter space corresponding to the physical world. Objects are distributed in this virtual space. Since the first recorded astronomical observation with a real telescope by Galileo more than 400 years ago, we have been constantly increasing the aperture of our telescopes. Only when photo plates were used, was the limited integration time of the naked eye (~0.1 s) surpassed. With a larger aperture and a longer integration time, new parts of the parameter space were revealed. Consequently, a large number of faint objects were detected.

Although the optical band, i.e., the frequency range of the visible light, is only a tiny fraction of the whole electromagnetic spectrum, it is one of the two atmospheric windows, i.e., the frequency range in which the Earth's atmosphere is nearly fully transparent to the electromagnetic wave. The other atmospheric window is part of the radio band, from about 5 MHz to 300 GHz. When Karl Jansky first detected the radio emission from the Galactic center, i.e., the center of our Milky Way galaxy in 1931, astronomers had access to this new part of the parameter space that was inaccessible with the naked eye and optical telescopes. These radio observations unfolded a new world.

Similar to the evolution of optical telescopes, the aperture of radio telescopes has been increasing. Receivers are also constantly upgraded with decreasing intrinsic noise, which is characterized by a parameter known as the system temperature $T_{sys}$. When a radio telescope with a larger aperture was built, or when an upgraded receiver was mounted, discoveries were made, since a new part of the parameter space became reachable.

The Five-hundred-meter Aperture Spherical radio Telescope (FAST) is a mega-science project in China.[1] It is more sensitive than other single-dish radio telescopes in the world, such as the Arecibo, Effelsberg, and Green Bank telescopes. The effective collecting area is about 1.6 times that of Arecibo.[2] FAST has a simple optics. It is currently possible to arrange a 19-beam L-band receiver, or even, in the future, an effective 100-beam receiver. With the 19-beam L-band receiver, FAST will cover a region of about 20′ × 20′ (note that the full Moon has an angular size of about 30′) with four pointings, or a range of declination of 22′ with one drift.[3,4] Also, thanks to its sophisticated feed support system, the area of the observable sky of FAST is about two times that of Arecibo, since the range of zenith angle of FAST is larger (see Table 3). Some important targets from the Arecibo sky, e.g., the M31 galaxy and the Orion molecular cloud, lie in the FAST sky. However, the observable sky of FAST is still much smaller than that of the fully steerable radio telescopes. The Galactic plane near the Galactic center is not accessible for FAST. This limits the potential pulsar discoveries, since the stellar density is highest in this region. The limited range of zenith angles of FAST also constrains its longest tracking time, which is crucial in the Very Long Baseline Interferometry (VLBI) observations.

Based on its characteristics, several scientific research projects have been designed for FAST, including pulsar search and pulsar timing, search for HI (neutral hydrogen) galaxies and HI mapping of the Milky Way, molecular line search, the search for extraterrestrial intelligence (SETI), joining the VLBI network, and several others. We give a brief introduction of FAST, present the scientific achievements of FAST, discuss the scientific prospects for FAST, and give a brief summary.

## Characteristics of FAST

The aperture of a traditional fully steerable radio telescope has been limited to about 100 m for half a century, since the 100-m Effelsberg telescope was built in the 1970s. The fully steerable radio telescopes in construction also have an aperture of around 100 m (e.g., the 110-m Qitai telescope and the 120-m Jingdong radio telescope). To push the limits, different designs or materials are needed. FAST is a telescope with a design that is different from a fully steerable telescope, sharing similar concepts with the 305-m Arecibo telescope. It is built in a karst depression to achieve a larger aperture and avoid the risk of flooding (Figure 1). Together with the active reflector and the flexible feed support system, these three main features give FAST its current status. The specifications of FAST are given in Table 1.

Incident radio waves first hit the reflector. A parabolic reflector is required because FAST is a prime focus telescope. As indicated by its name, the reflector of FAST is a spherical cap. Therefore, part of the reflector dynamically forms a paraboloid during observations. The aperture of this instant paraboloid is 300 m.

The focused radio wave from the instant paraboloid is collected by the receiver on the focal plane. Unlike the fully steerable telescopes, there is no rigid connection between the receiver and the reflector. The optics of FAST relies on the accurate measurement and control of the shape of the reflector and the position of the receiver. The efficiency is higher than 60% within the zenith angle of 26.4°. The pointing accuracy is better than 16″.[2]

FAST reveals a new part of parameter space comprising several parameters: e.g., raw sensitivity, time resolution, frequency resolution, sky coverage, and number of beams. Of these parameters, the key is raw sensitivity, defined as the ratio of the effective collecting area to the system temperature, $A_{eff}/T_{sys}$. With the 19-beam L-band receiver, which has a frequency range of 1.05–1.45 GHz, FAST has a maximum raw sensitivity of about 2,600 m$^2$/K.

The time resolution of the current pulsar observation of FAST is about 50 μs. The frequency resolution for SETI observations can reach 5 Hz.[2] Combined with the high sensitivity of FAST, a new part of parameter space can be revealed: we can study short timescale phenomena, such as single pulses of pulsars and fast radio bursts (FRBs). With baseband data recording, higher time and spectral resolution are also possible, however, this requires a large amount of storage and computing resources. One should note that the time resolution $\Delta t$ and the frequency resolution $\Delta \nu$ fulfill the uncertainty relation $\Delta t \cdot \Delta \nu \sim 1$. It is not possible to simultaneously achieve a high time resolution and a high frequency resolution.

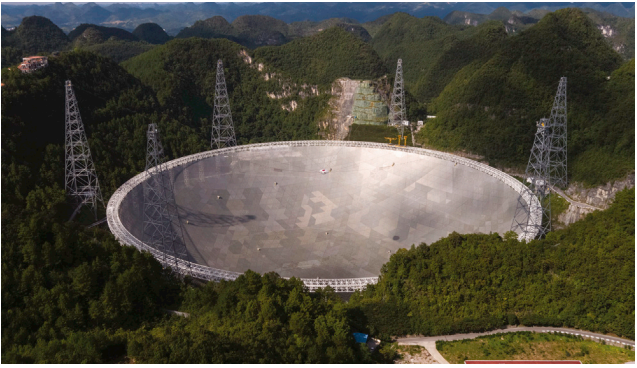

Figure 1. The Overview of FAST FAST was built in a karst depression in Guizhou Province, southwest China.

Table 2. FAST Receivers

| No. | Frequency Range (GHz) | Remarks |
|---|---|---|
| 1 | 0.07–0.14 | |
| 2 | 0.14–0.28 | |
| 3 | 0.27–1.62 | ultra-wideband receiver |
| 4 | 0.56–1.12 | |
| 5 | 1.15–1.72 | |
| 6 | 1.04–1.45 | 19-beam L-band receiver |
| 7 | 2.00–3.00 | |

The maximum zenith angle of FAST is 40°, so the observable sky of FAST covers a declination range of −14.4° to 65.6°. The total solid angle is 2.32π, about 58% of the whole sky (4π).

FAST covers the frequency range of 70 MHz to 3 GHz, with 7 sets of receivers (Table 2). The current L-band multibeam receiver of FAST has 19 beams, with the beams arranged in a hexagonal configuration.[3]

For comparison, we have listed some key parameters of several large radio telescopes in Table 3. FAST has a relatively high gain and low system temperature, but the range of the zenith angle is limited. As a single-dish telescope, the beam size of FAST is also larger than the telescope arrays. In the future, the gain, and system temperature of FAST can be improved with more accurate control of the shape of the reflector and more sophisticated receivers. The maximum zenith angle can also be extended in the future.

### Scientific Achievements of FAST

As a large-aperture radio telescope, the high sensitivity, high time resolution, and high frequency resolution of FAST have been made best use of to detect weak signals or short timescale signals. The observations of pulsars, FRBs, and SETI research fit well into this regime. There have been several scientific achievements using FAST in these areas.

Pulsars are recognized as rotating neutron stars. They are celestial laboratories of all four fundamental interactions. Worldwide, about 3,000 pulsars have been discovered. Complex and rich phenomena have been observed in these pulsars. The profile and polarization of the pulses give hints to the structure of the emission regions in the magnetosphere of pulsars. The dynamic spectra of pulsars have also been used to study the scintillation caused by interstellar plasma. The time of arrival (ToA) of the pulses also contains abundant information. Pulsar timing, the study of ToA of pulsars, has helped us to get indirect evidence of gravitational radiation in binary pulsar systems. We also expect that the imprint of the $10^{-9}$ Hz gravitational wave has been hidden in the timing noise. It is possible to detect the gravitational background using the pulsar timing array (PTA).[5]

Since the discovery of the first pulsar,[6] more than 240 pulsars have been discovered by FAST. Among these pulsars, about 30% are millisecond pulsars. There are also pulsars of special kinds, e.g., pulsars with nulling. Pulsars are also confirmed (in M13),[7] or discovered (in M92),[8] in the tracking observations of globular clusters. Both pulsars lie in the binary systems, showing the signature of the eclipse by their companions (Figure 2). Follow-up observations of these newly discovered pulsars are currently on-going. The radiation mechanism of pulsars has been studied by analyzing the structure of single pulses.[9,10] The scintillation of interstellar plasma has also been studied using pulsar observations,[11] and characteristic scintillation arcs have been detected (Figure 3).

FRBs are a short timescale phenomenon discovered in recent years.[12] Generally, they show pulses of ~1 ms in width. We still know little about the nature of FRBs, but the detections of them are rapidly increasing, and there have been suggestions to use FRBs to study the intergalactic medium and cosmic turbulence.[13] Observations of FRBs have become an important time domain science for FAST. A new FRB (FRB 181123) has been discovered in the drift scan survey of FAST.[14] This source shows a downdrift pattern that is common in repeating FRBs (Figure 4), but there has been no evidence that this FRB is repeating. Compared with the famous FRB machine, CHIME (https://chime-experiment.ca/en), the FRB discovery efficiency of FAST is not high. About one FRB will be discovered in FAST observations of about 1,000 h. So FAST searches for FRBs should be done commensally with other surveys. On the other hand, FAST is good at searching for repeating pulses from known FRBs. FAST observed the repeating FRB, FRB 121102. A set of pulses from this source has

Table 1. FAST Specifications

| Specifications | Value |
|---|---|
| Illumination aperture | 300 m |
| Declination (Dec) range | −14.4° to 65.6° |
| Frequency range | 70 MHz to 3 GHz |
| Beam size (L-band) | ~2.9′ |
| Maximum raw sensitivity (L-band) | ~2,600 m²/K |
| Maximum time resolution | 50 μs |
| Maximum frequency resolution | 5 Hz |

Table 3. Comparison of FAST with Other Large Radio Telescopes

| Telescopes | Gain | $T_{sys}$ | Beam Size | Zenith Range | Remarks |
|---|---|---|---|---|---|
| | (at 1.4 GHz, K/Jy) | (at 1.4 GHz, K) | (at 1.4 GHz) | | |
| FAST | 16.1 | 18 | 2.9′ | 0°–40° | 1 |
| Arecibo | 10 | 30 | 3.5′ | 1.06°–19.69° | 2 |
| GBT | 2 | 18 | 8.8′ | 0°–85° | 3 |
| Effelsberg | 1.55 | 23 | 9′ | 1°–81.9° | 4 |

Remarks: (1) the gain and system temperature is the value for a zenith angle below 26.4°[2]; (2) http://www.naic.edu/science/generalinfo_set.htm; (3) https://greenbankobservatory.org/science/telescopes/gbt/; (4) https://www.mpifr-bonn.mpg.de/effelsberg/astronomers.

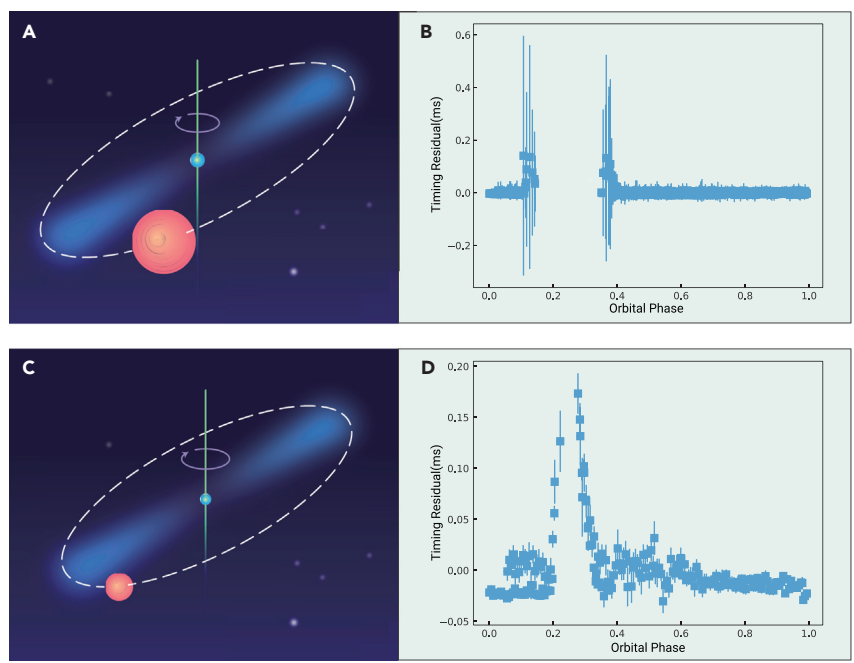

**Figure 2. The Sketch and the Timing Results of M92A and M13E** (A) A cartoon to show the binary system M92A.
(B) The timing residual of M92A at different orbital phases. The eclipse between the orbital phase of 0.1 and 0.4 is evident. The large deviation at the two edges is due to the atmosphere of the companion star. Reproduced from Pan et al.[8] with permission.
(C) A cartoon to show the binary system M13E.
(D) The timing residual and dispersion measure of M13E at different orbital phases. The eclipse between the orbital phase of 0.2 and 0.3 is evident. The large deviation at the two edges is due to the atmosphere of the companion star. Reproduced from Wang et al.[7] with permission.

been detected by FAST.[15] This is currently the largest pulse set of an FRB. Recently, a repeating FRB, 180301, was confirmed by FAST. The polarization observation also indicated the magnetospheric origin of the FRB emission for the first time.[16] Observations of the Galactic magnetar SGR J1935+2145 also give a stringent constraint on the FRB-SGR burst association.[17]

Another time domain study using FAST is of SETI in the radio band, i.e., radio SETI. This is a serious scientific goal of FAST from the design phase and has been included in current surveys.[18] Recently, the first SETI observation using FAST was reported,[19] marking FAST as a powerful machine in SETI signal searching.

For spectral observations, HI mapping of the Milky Way and searching for HI galaxies have been done commensally with pulsar searches in a drift scan survey of a small part of the sky. The high-frequency calibration technique makes commensal observation of pulsar and spectrum possible, since the pulsar observations are not affected by this noise injection for spectral calibration.[3]

The famous molecular line sources Orion BN/KL and TMC1 have been observed to search for molecular lines in the frequency range of 1.05–1.45 GHz. The search for a new molecular line has not been successful, but many recombination lines of hydrogen and carbon have been detected (Shi Hui, personal communication). HI gas has also been detected in galaxies beyond the Milky Way with long time integration.[20]

### The Prospects of FAST Science

As mentioned above, the key scientific goals of FAST include pulsar search and pulsar timing, search for HI galaxies and HI mapping of the Milky Way, molecular line search, SETI, and to join up with the VLBI network.

Future observations with FAST will be a further exploration into the relevant parameter space. The high sensitivity, high time resolution, and high frequency resolution of FAST have been fully utilized in current observations. Future studies in both time domain and frequency domain will also rely on long-term observation of the sky area accessible by FAST with an even higher survey speed (defined as the sky area surveyed in unit time) provided by the phased array feed receiver, a new kind of receiver with more effective beams. A systematic survey of the FAST sky has just started. As the observed sky area increases, there will be increasing discoveries of objects (and phenomena), such as pulsars, FRBs, HI galaxies, interstellar molecules, and megamasers (see Table 4).

Despite the familiar parameter space that we can conceive, there are also possible new parts of the parameter space built up by a new combination of parameters. As an example, the time variation of spectral lines is still lacking in meticulous studies. These new parts of the parameter space can be attributed to the notion of "human bandwidth,"[21] emphasizing that the new parameter space would be found in someone's idea.

*Time Domain Studies.* With regard to searching for pulsars, FAST is constantly detecting new candidates. Currently, there are several ongoing pulsar surveys using FAST, covering the Galactic plane, the Galactic halo, globular clusters, and the M31 galaxy; it is estimated that more than 1,000 pulsars will be discovered. In these new pulsars, there could be several binary systems, especially extreme relativistic binary systems, e.g., double pulsars, even pulsar-black hole binaries. These binary systems could provide a better test of gravity theories. More pulsars with nulling, subpulse drift, and mode change could also be discovered, contributing to the study of the radiation physics of pulsars.[22] There could also be more pulsars with glitch, i.e., a sudden change of the rotating period. All of these studies rely on the systematic observation of the timing of the pulsars after their discovery.

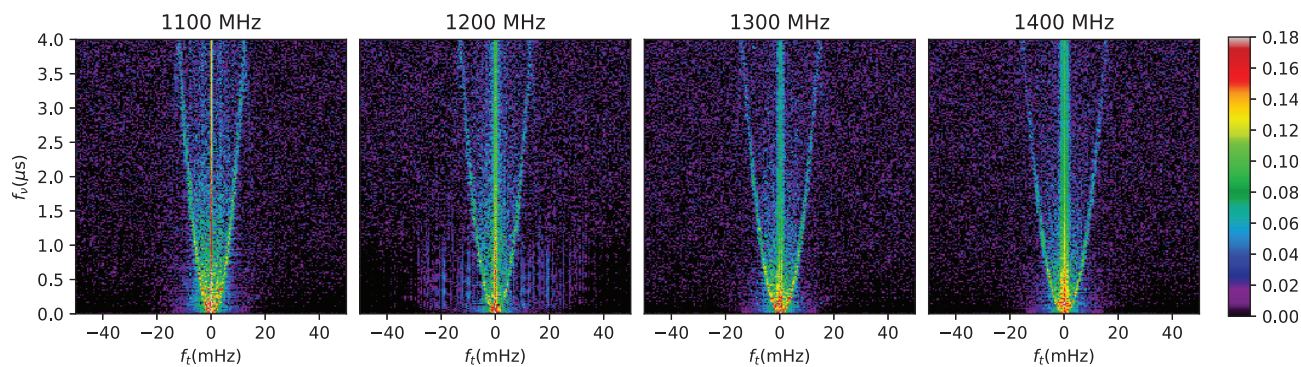

**Figure 3. The Secondary Spectra of PSR B1929+10 from Four Subbands** The colors show the relative strengths. Units are arbitrary. From top to bottom, the subband central frequencies are 1,100, 1,200, 1,300, and 1,400 MHz, respectively. Reproduced from Yao et al.[11] with permission.

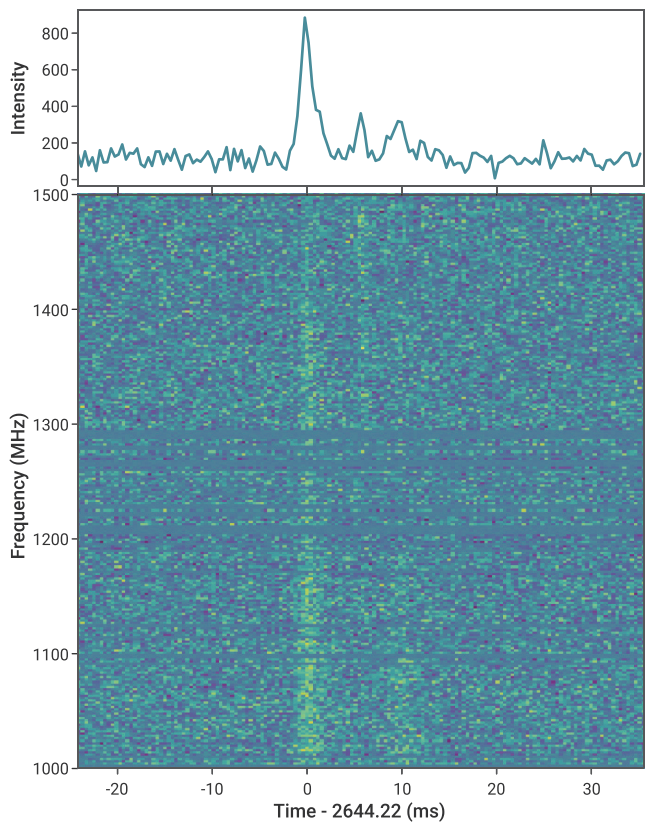

Figure 4. The Pulse Profile (Top) and the Dynamic Spectrum (Bottom) of FRB 181123 Reproduced from Zhu et al.[14] with permission.

Pulsar timing is a journey that requires patience, along one dimension of the space parameter: time. Newly discovered pulsars need at least 1 year for their coordinates be accurately obtained by timing observations. The test of gravity theory with the observation of the orbital evolution of a binary pulsar usually takes several years. The collection of a large sample of glitches takes a long time. Searches of the nano-Hertz gravitational wave with PTA also need a time baseline of at least 3 to 5 years.

Searches for FRBs and the SETI signal should be done commensally with pulsar searches. This strategy effectively increases the observation time. As the observation proceeds, more FRBs could be detected and a more stringent constraint on the SETI signal will be given. Based on a larger set of FRB pulses, the radiation mechanism and even the potential period can be revealed.

Besides these well-known targets, it is also proposed to search for radio pulses produced by the impact of high energy cosmic rays on the surface of the Moon.[23]

**Frequency Domain Studies.** Studies of atomic hydrogen gas in our Milky Way and other galaxies using FAST rely on large-scale surveys and long-term observations.[3] HI mapping of the Milky Way will become more complete as the survey proceeds. Finally, a full HI map of the whole sky observable by FAST will be produced.

Table 4. Some of the FAST Science Targets and Their Requirements

| Science Targets | Requirements |
| --- | --- |
| Pulsars | 1, 2, 4 |
| FRBs | 1, 2, 4 |
| SETI | 1, 2, 3, 4 |
| HI | 1, 3, 4 |
| Molecular lines | 1, 3, 4 |
| Megamasers | 1, 3, 4 |
| Scintillation | 1, 2, 3, 4 |

1, Raw sensitivity; 2, time resolution; 3, frequency resolution; 4, observation time.

With regard to the drift scan survey, it is expected that it will detect $6.5 \times 10^5$ HI galaxies with a median redshift of 0.07, with a mass threshold of $10^{9.5} M_\odot$.[4] As well as the blind survey using the drift scan mode, the targeted HI surveys of the galaxies selected in other bands will also be important work for FAST.

As well as the 1,420 MHz HI line, there are also other lines in the FAST frequency range, e.g., the recombination lines of hydrogen, helium, and carbon, OH lines (1,612, 1,665, 1,667, and 1,720 MHz), CH lines (702, 704, 722, and 725 MHz), and $CH_3OH$ lines (834 MHz) (http://www.cv.nrao.edu/php/splat/). The farthest 1,667 MHz OH megamaser currently known has a redshift of 0.265. FAST can detect an OH megamaser out to $z \sim 2$,[24] and will also continue to search for new molecular lines. A new ultra-wideband receiver covering roughly 500 MHz to 3 GHz will be built in the near future (Liu Hongfei, personal communication). It will become easier for line search when this new ultra-wideband receiver is ready. It is also proposed to constrain the variation of the basic physical constants with molecular line observations.[25] The cosmic acceleration can also be directly measured by observation of HI absorption systems.[26]

The continuum survey will also be done commensally with the HI survey. For point sources, FAST is not as good as an interferometer, such as VLA (https://science.nrao.edu/facilities/vla), MeerKAT (https://www.sarao.ac.za/gallery/meerkat/), and ASKAP (https://www.atnf.csiro.au/projects/askap/index.html). But FAST will be useful in mapping extended structures such as supernova remnants.

**Other Works.** Unlike traditional studies with a division between the time and frequency domains, some studies sahould become a combined time-frequency analysis.

The study of scintillation (see Table 4) is a typical time-frequency analysis. Based on the preliminary results, it has been shown that FAST is quite suited for scintillation studies.[11] Since the raw sensitivity of FAST is high, it is possible to detect structures in the interstellar medium that are invisible to other smaller telescopes.

It is also proposed to study radio emissions from the interaction between the erupted plasma from a star and the magnetic field of the planets around the stars.[27] This would be an analysis of the flux distribution on the time (orbital phase)-frequency plane.

### Conclusion

FAST opens up a new part of the parameter space in the exploration of the universe. The high sensitivity has been made best use of in time domain observations, with fruitful results, e.g., the discovery of more than 240 pulsars, the discovery of FRBs, and the collection of the largest set of pulses from repeating FRBs.

Future scientific exploration using FAST will be a journey into new parts of parameter space. Most scientific goals, such as the systematic surveys of pulsars and galaxies, the mapping of HI of the Milky Way, and the timing of pulsars, need a time baseline of several years. This will be a journey demanding patience and willpower.

As a new dimension of the parameter space, the human bandwidth is notable. More users are encouraged to observe using FAST and to contribute ideas. The FAST team has called for proposals from astronomers working in China in 2020. The calling for worldwide proposals will come in 2021. With more proposals, the human bandwidth of FAST will increase and a larger part of the parameter space will be accessible.

### ACKNOWLEDGMENTS

This work is supported by the National Key R&D Program of China no. 2018YFE0202900. L. Qian is supported by the Youth Innovation Promotion Association of CAS (ID 2018075). R. Yao is supported by the Youth Innovation Promotion Association of CAS (ID 2017080). J.H. Sun is supported by the Youth Innovation Promotion Association of CAS (ID 2016059). J.L. Xu is supported by the Youth Innovation Promotion Association of CAS (ID 2019058). Z.C. Pan is supported by the CAS "Light of West China" Program. P. Jiang is supported by the Youth Innovation Promotion Association of CAS (ID 2013039). L.Qian is indebted to the helpful discussion with Dr. Chengjin Jin , Dr. Di Li , Dr. Ningyu Tang, Dr. Hongfei Liu, Dr. Youling Yue, Dr. Hui Shi, Dr. Liang Dong, and Dr. Yonghua Xu. Leas Contact Website https://fast.bao.ac.cn/.